# Engineering a sustainable world by enhancing the scope of systems of systems engineering and mastering dynamics

Rasmus Adler
Fraunhofer IESE
67663 Kaiserslautern, Germany
rasmus.adler@iese.fraunhofer.de

Frank Elberzhager
Fraunhofer IESE
67663 Kaiserslautern, Germany
Frank.Elberzhager@iese.fraunhofer.de

Florian Balduf
Fraunhofer IESE
67663 Kaiserslautern, Germany
florian.balduf@iese.fraunhofer.de

**Abstract**. Engineering a sustainable world requires to consider various systems that interact with each other. These systems include ecological systems, economical systems, social systems and technical systems. They are loosely coupled, geographically distributed, evolve permanently and generate emergent behavior. As these are characteristics of systems of systems (SoS), we discuss the engineering of a sustainable world from a SoS engineering perspective. We studied SoS engineering in context of a research project, which aims at political recommendations and a research roadmap for engineering dynamic SoS. The project included an exhaustive literature review, interviews and workshops with representatives from industry and academia from different application domains. Based on these results and observations, we will discuss how suitable the current state-of-the-art in SoS engineering is in order to engineer sustainability. Sustainability was a major driver for SoS engineering in all domains, but we argue that the current scope of SoS engineering is too limited in order to engineer sustainability. Further, we argue that mastering dynamics in this larger scope is essential to engineer sustainability and that this is accompanied by dynamic adaptation of technological SoS.

## Introduction

Sustainability refers to the societal goal that broadly aims for humans to safely co-exist on earth within planetary boundaries, and thus, for a long time. Currently, planetary boundaries are violated. This has negative impact on the boundaries and makes it urgent to reach a situation where the boundaries are no longer violated. Instead of hoping that the traditional evolution of existing systems will bring us somehow to this situation, a more promising and reasonable approach is to systematically steer into this direction, that is, to engineer sustainability. However, engineering a sustainable world is a huge challenge. It is even hard to define more clearly what sustainability actually means. A prominent approach is given by the sustainability development goals (SDGs). This approach does not decompose the big challenge of achieving sustainability down into independently solvable sub-problems. Ecological systems, economical systems, social systems, technical systems and other kinds of systems have complex interdependencies between each other and to the 17 SDGs [1] and the related 169 sustainability targets refining the SGDs. If there are several options to influence some systems, then it is typically challenging to choose the one which has the biggest contribution to the 169 sustainability targets. Interdisciplinary science and systems engineering are key to cope with this challenge, because the different kinds of systems require to involve different kinds of disciplines.

Historically, some disciplines are deeply rooted in systems engineering. In interviews that we conducted in context of the project DynaSoS [2][3], interviewees saw systems engineering as a transdisciplinary approach that brings together mechanics, electronics and informatics in order to engineer technical systems. Sociotechnical aspects were also considered but mainly in order to derive requirements for the technical systems. Economic aspects were considered in context of the business models. Regulations were merely considered as a constraint. Following this narrow scope, engineering sustainability is limited to the fulfillment of existing regulations, sustainable business models, and sustainable financing. Engineering in this narrow scope with fixed governance and regulation cannot steer a transformation. One can only implement transformations according to a given direction. If the direction is wrong, then systems that are not sustainable are strengthened. This hinders necessary fundamental changes. Accordingly, we argue that this common perspective on systems engineering is too narrow for engineering inherent sustainability. We propose to **research in the direction of a larger engineering scope** that focuses more on non-technical systems such as ecological, economic and social systems. In this scope, possible changes are not limited to technical systems. For instance, regulation is not considered as constraint but as an adjustable influence factor. This is already the case in regulatory science which aims at scientific foundations for regulations. It focuses traditionally on industries involving health or safety. We argue that this will and should change considering the number of recent and upcoming regulations related to sustainability. Systems engineering needs to involve regulatory science but also various other scientific disciplines like economic science in order to steer towards sustainability. One could argue that the different disciplines have to work together anyway to achieve sustainability and systems engineering provides the means for the necessary transdisciplinary communication and documentation.

A second proposal that we present in this paper is to **research the dynamics in SoS**. We observed in course of the DynaSoS project that the visions in various domains like mobility or energy refer to system transformations that increase dynamics. For instance, on-demand shared mobility implies dynamic adaptation to mobility demands and possibly dynamic pricing according to current particle pollution status or demand/offer. Another example are smart grids, which adapt to demands and use dynamic pricing as a means to influence the demands and to adapt to current weather conditions. As illustrated by the examples, dynamism can contribute to efficiency, which is one of the main strategies for sustainability. Further, they illustrate that it can contribute to sufficiency, which is another very important strategy that complements efficiency and addresses related rebound effects. Various domains engage similar approaches to address this dynamic aspect. The role of systems engineering here is to foster cross-fertilization and to avoid that some domains reinvent the wheel. Further, it should take care that solutions form different domains are known and compatible. This is important as a SoS grows and can involve systems from different domains.

In the remainder of this article, we will further justify and refine our two proposals for further research. To this end, we will first share our observations in the DynaSoS project. Next, we will reflect on these observations and discuss the two proposals. Finally, we will summarize our thoughts and draw conclusions with respect to the engineering of sustainability.

## Observations

In the course of the DynaSoS project, a team of 38 experts performed an exhaustive literature review and conducted a substantial amount of interviews and workshops with representatives from industry and academia from different domains. In order to collect research challenges and derive a related research roadmap for engineering SoS. We will first summarize which findings from interviews and workshops are relevant for engineering sustainability. Afterwards, we will summarize relevant findings from the literature review.

### *Interviews and Workshops*

Before we share our observations, we give some relevant background information about the interviewees and workshop participants.

**Background information.** The group that organized the interviews and workshops were domain experts or system engineering experts. Accordingly, interviewees and workshops focused either on domains or on systems engineering in general. The systems engineering group involved members of the German Chapter of INCOSE "Gesellschaft für Systems Engineering" and many prominent systems engineering experts in Germany. The results have been presented in [4] and discussed at the day of systems engineering in November 2022. Similarly, the domain experts involved relevant organizations and experts from their domain. For instance, the central network to advance digital transformation in production in Germany, Plattform Industrie 4.0 [5] was involved for the manufacturing domain.

**Findings.** In all domains, **sustainability was a major driver** for the (digital) transformation of existing SoS. Domain experts were often not familiar with the term SoS but their visions referred to transformations of SoS or the introduction of novel SoS. This lack of familiarity with these concepts exhibits ones more the need of cross-fertilization between domains. Most of them were also not aware of the following five SoS characteristics [6], but many challenges that they described were related to these characteristics.

1) *Operational independence:* In a SoS, Constituent Systems (CS) operate independently of the SoS and other systems.

2) *Managerial independence:* The CS in a SoS are managed independently and their owner/managers may be evolving the systems to meet their own needs.

3) *Geographical distribution:* In many cases, constituent systems in a SoS are geographically distributed, although many view this as a less significant or secondary characteristic of SoS.

4) *Evolutionary development:* SoS development is based on developments in the CS. These developments may take place asynchronously based on the independent development processes of the CS. This means that the SoS will evolve incrementally rather the be 'delivered' as normally envisioned in a single system development or acquisition.

5) *Emergent Behavior:* Emergent system behavior can be viewed as a consequence of the interactions and relationships between system elements rather than the behavior of individual elements. It emerges from a combination of the behavior and properties of the system elements and the systems structure or allowable interactions between the elements, and may be triggered or influenced by a stimulus from the systems environment

In many cases, the **first and major challenges were not technical** and related to managerial independence. Owner/managers evolve the systems to meet their own needs. These own needs were not in line with overall objectives. However, many other actors are relevant, because they influence the needs of the owner and manager. Accordingly, there is a bigger picture that explains why overall objectives are not fulfilled. Actors simply do have not the right incentives, are hindered by current regulation or they do not have sufficient financial opportunities. Domain experts mentioned such non-technical issues often as their first big hurdle. Particularly, mobility experts mentioned that shared multi-modal mobility is rather hindered by non-technical systemic issues than by technical issues such as the development of safe autonomous shuttles. Some challenges could be addressed with follow-up research projects but the DynaSoS project aims at a research roadmap addressing software-related technical issues. This exemplifies the kind of systemic non-technical issues that we observed. To address such systemic non-technical issues, we propose to **broaden the scope of systems engineering** and to apply systems thinking in order to understand the multi-actor network in the considered sociotechnical system or sociotechnical regime as it is called in [7].

Considering the currently dominant narrow scope of systems engineering and the technical focus, we observed that the visions from the domain experts came along with an **increase of dynamics and**

**runtime adaption**. We already exemplified this with the adaptive behavior of envisioned smart grids and mobility systems that consider currently available mobility resources and current demands. Flexibility based on dynamics was also very prominent in the smart manufacturing domain. Elements in a supply chain are transformed so that they offer flexibility. A prominent example was the adaptable factory. Based on this flexibility, the supply chains can be transformed in a resilient supply network that dynamically adapts in order to withstand various disturbances. Related visions exist in the agriculture domain. Automation in manufacturing differs from automation in farming but the concepts are the same. Also, the visions hint to the same direction, because a food supply network is a special case of flexible supply network. Farming makes a food supply network special, because the time of growing limits the flexibility and the related runtime adaptation. The motivation for runtime adaption is not affected and the same concepts are applicable for achieving it.

In contrast to the domain experts, the systems engineering experts knew the term SoS, but some of them were confused by the attribute "dynamic". This triggered us to think about related terms like "dynamical" or "adaptive". We are not referring to the dynamical systems in terms of mathematics or physics. The formal foundations of dynamical systems theory can contribute to the engineering of dynamical SoS, but we do not want to limit the problem space to those problems that can be addressed with theory. The notion of "adaptive" as it is considered in Software Engineering for Adaptive and Self-Managing Systems (SEAMS) and related Dagstuhl seminars [8] is very close to our notion of "dynamic". A difference is that the "adaptive" relates to the behavior of technical systems whereas our "dynamic" also refers to the behavior of other systems.

The systems engineering experts mentioned many challenges with respect to the engineering of SoS. They had two perspectives when thinking about SoS and related challenges. The first one was that SoS become more complex due to technological innovations based on connectivity, digital twins, AI and so on. The second one was that technological innovations like digital twins and AI can be used to deal with the complexity. In this context, many highlighted the need for automation of systems engineering tasks. None of them mentioned out of themselves that the scope of engineering might be too narrow. Also, none of them mentioned that increasing dynamism needs to be addressed.

### *Literature study*

The literature review in the DynaSoS project considered various research fields. Most related to our proposals was literature from SoS engineering, complexity science, sustainability science and safety science. In the following, we present some findings related to these fields.

**SoS engineering.** SoS engineering has a long history and goes at least back to 1998 where the popular paper "Architecting Systems of Systems" [9] was published. SoS engineering evolved since this time. Literature stimulated this by proposals for further development. This was also so in the last decade. In 2007, a list of ten research themes has been proposed in [10] based on a workshop with 13 participants. In 2013, 78 research elements were grouped into twelve major themes and prioritized by a community of 38 experts [11]. In 2016, a paper [12] was published where 32 primary studies (31 between 2005 and 2015, plus one study from 1994) have been analyzed in order to identify initiatives, trends, and challenges in SoS development. In 2018, challenges related to requirements engineering for SoS were presented in [13] and a related list of ten research themes was presented. In 2020, a paper [14] focused on SoS architecture and considered publications over the last ten years. In 2022, a paper [15] considered 41 studies (between 2006 and 2021) in order to extract challenges related to architecture evaluation. Particularly, worth mentioning is the INCOSE's Vision 2035 [26] (and its earlier versions: Vision 2020 and Vision 2025), which considered SoS aspects.

Considering these and other related work, we make the following two main observations. First, sustainability is increasingly mentioned as motivation for SoS engineering. Second, there is a tendency to consider larger SoS and to consider more and more application domains. In the beginning, the focus was on defense, national security, military and aerospace. All these domains have in common that there are big (publicly funded) organizations taking care of these SoS. Further, transportation and energy came to the fore very early. Today, the word "smart" indicates in various domains that SoS engineering is needed. Smart city, smart manufacturing, and smart farming are prominent examples. These smart systems are increasingly connected to each other. This interconnectedness has reached a global scale and introduces a tremendous complexity. INCOSE's Vision 2035 has a dedicated chapter addressing the "global context of systems engineering" where it mentions that "complexity results from the increasing software and data content, increasing systems interconnectedness, competing stakeholder expectations, and the many other social, economic, regulatory, and political considerations that must be addressed when designing systems in a SoS context". Further, it mentions that "systems engineering is inherently trans-disciplinary, and must include representation and considerations from each discipline and each affected stakeholder". All this is in line with our proposal to enlarge the scope of engineering even though some disciplines that we see as relevant for engineering a sustainable world are not in the focus. It also motivates to have a closer look on the evolution of complexity science.

**Complexity science.** Complexity science is a huge field that has also a very long history. The map of complexity science (cf. Figure 1 and [16]) by Brian Castellani & Lasse Gerrits provides a reasonable overview.

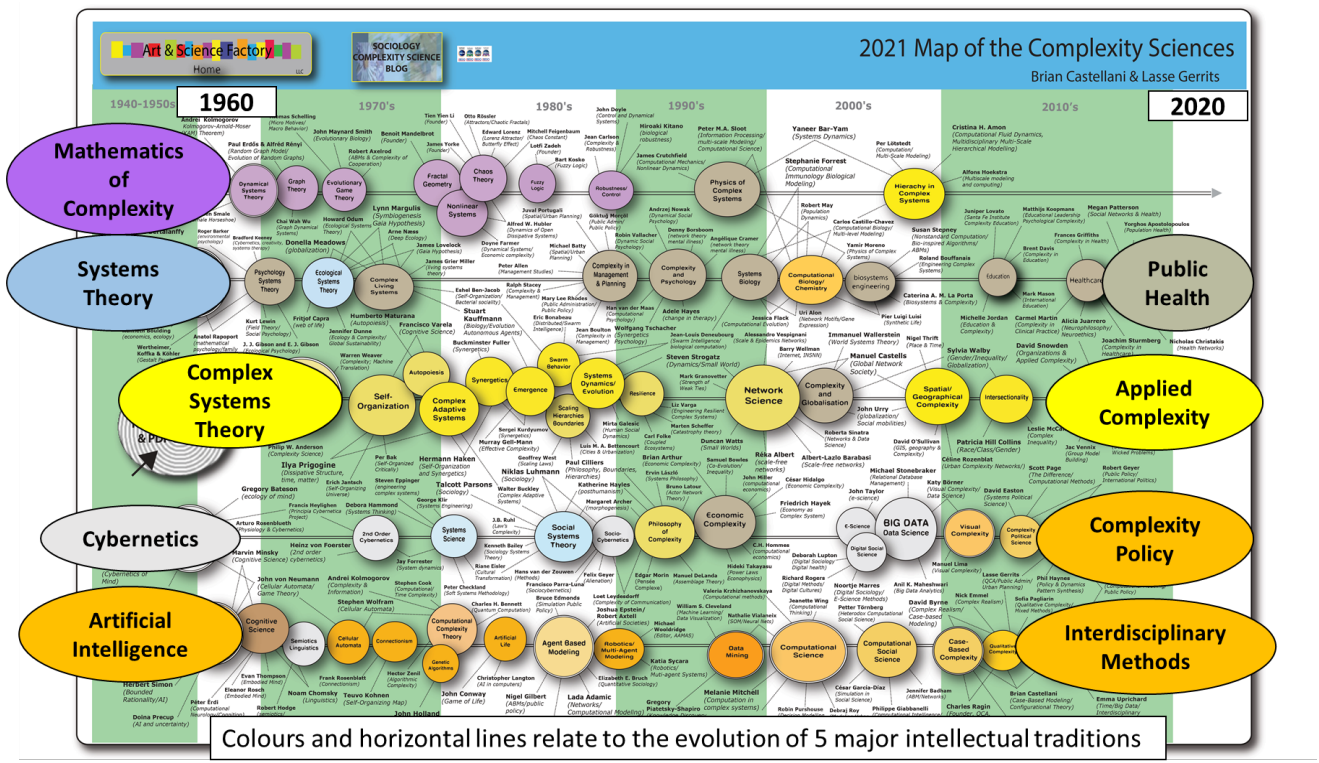

Figure 1 – 2021 Map of Complexity Science

According to the structure of the map, complexity sciences goes back at least to the 1960. It is rooted in mathematics of complexity, (complex) systems theory, cybernetics and artificial intelligence. The roots have developed towards research areas called "public health", "applied complexity", "complexity policy" and "interdisciplinary methods". These headings and the corresponding references to publications and organizations relate to our proposal of a broader engineering scope. Some of the referenced work also creates a link to the dynamic aspect that we identified as an important characteristic. For instance, a webinar [17] from the organization corresponding to area of "complexity policy" explains how system dynamics can support policy evaluation.

**Sustainability science.** UNESCO describes in sustainability science in [18] as 1) science about sustainability, to understand how complex physical, biological and social systems function, and 2) science for sustainability, to support sustainable policies and positive social transformations. Other descriptions outline a similarly large and open field. In this field, we searched for work that is related to our proposals and helps to better define what we mean by "a lager engineering scope" and "dynamic". In the following, we mention some of the work that we have found.

The working paper "Dynamic Systems and the Challenge of Sustainability" [19] provides some descriptions that fit to our proposals. Our proposal to enlarge the scope of engineering means that we come from a technical focus and enlarge it with respect to other systems and functions. It does not mean that we lose sight of technology. For instance, the UNESCO description above does not even mention technology. Humanity would not have been able to create such huge environmental impact without technology. The role of technological systems should thus be considered in the larger engineering scope. The working paper nicely describes this role: "We consider interactions of biological/ecological and social/economic/political systems in the developing world, and the mediating role of technology in altering and being altered by natural and social-political systems". It also provides a description of dynamics that is in line with the kind of dynamics that needs to be considered in the larger engineering scope: "'Dynamics' refers to the patterns of complexity, interaction (and associated pathways) observed in the behavior over time of social, technological and environmental systems". In the narrow scope that focuses on technology, our notion of dynamics is closer to the one of adaptive systems as we discussed it previously.

Sustainability theories such as the resilience theory [20] or the multi-level perspective [7] provide terms and models to describe our proposals. In this paper, we will relate our proposal to enlarge the engineering scope to the multi-level perspective. The latter describes how current technological regimes tend to support incremental evolution of current technology and to strengthen the current regime. The stronger the current regime, the less likely other technologies will enter the regime. The regime includes producer networks, suppliers, financial networks, research networks, user groups, societal groups and public authorities. In addition to the term regime, the multi-level perspective refers to niches and landscapes. Niches are the places where innovations can grow, because they are protected from some forces like economic forces that exist in the regime. The landscape defines deep structural trends for the regime such as wars, emigration, cultural and normative values, and environmental problems. Traditional SoS engineering focuses on the regime, accepts it and tries to further develop the existing SoS in incremental steps. Our idea of the enlarged scope of SoS engineering is to take less for granted when engineering within the regime and to think of opportunities to influence the landscape.

Enlarging the scope of engineering in this way introduces complexity. This raises the questions which modeling and analyses approaches from complexity are suitable to deal with this complexity. Related work in this regard is presented in [21]. It provides a framework that is based on complexity theory and supports the modelling of transitions to sustainability following sustainability theories such as the multi-level perspective.

**Safety science.** Safety science also inspired us to propose a larger engineering scope. In safety science and related regulatory science, regulation is scrutinized. Further, systems thinking is well established in safety science. The popular book with the title "Engineering a Safer World: Systems Thinking Applied to Safety" [22] demonstrates with many examples how systems thinking can identify systemic failures. It proposes hierarchical control loops as means to engineer safety at various levels. One example shows relationships between congress, legislature, regulatory agencies, industry associations, unions, insurance companies, and courts. In the last years, safety was heavily studied from such a systemic perspective. For instance, the international collaboration community Engineering X launched in 2019 a £5 million five-year mission, Safer Complex Systems, to enhance the safety of complex infrastructure systems globally. The homepage mentions that "food and water supply, healthcare, education, housing, transportation and communications, is made up of complex systems that are highly interconnected and interdependent on one another. As the devastating COVID-19 pandemic has demonstrated, when one complex infrastructure system fails, many other complex systems are also affected, which can have catastrophic consequences for people's lives" [23]. In this regard, safety science already considers a very large engineering scope. The scope is limited to safety of the current population. In contrast to that sustainability takes also safety of future generations into account. In this regard, sustainability enlarges safety. Accordingly, it is nearby to investigate if the engineering approach for safety can be enlarged to sustainability. Some initiatives and proposals already hint to this idea, such as the working group "safety of the environment" of the safety-critical systems club [24] or the paper "global warming and system safety" [25].

## Reflection and Proposals

In this section, we will reflect on the presented observations and discuss the two proposed research directions. First, we discuss the direction towards a larger engineering scope and then the direction towards dynamics.

### *Research addressing a larger scope of SoS engineering*

We observed in the interviews and workshops that many sustainability challenges cannot be addressed by the currently limited scope of SoS engineering. This is because systemic failures due to inadequate governance cannot be addressed. Further, we observed that the idea of a larger engineering scope is not spread in the (German) systems engineering community. Law, regulations, and policies are a fixed context constraint. This is also reflected in the SoS Standard ISO/IEC/IEEE 21840 from 2019.

On the other hand, sustainability science, complexity science, and safety science already follow the idea to engineer proper governance. In doing so, they apply transdisciplinary systems thinking, which is the essence of systems engineering. Consequently, we propose that SoS engineering should be more involved in these activities.

In line with this proposal, INCOSE's Vision 2035 [26] mentions on page 8 that "complexity increases with systems scope expansion due to proliferation of interfaces and governance mechanisms". Accordingly, it mentions on page 29 for the future of systems engineering that "systems engineering will be broadly recognized by governments and industry as a high value contributor, resulting in a growing demand for systems engineering education and skills." The executive summary for the vision ends with the statement that "realizing this vision will require collaboration and leadership across industries, academia, and governments to meet these challenges and implement the high-level roadmaps outlined in Chapter 4."

This raises the question why our interviewees did not mention this governance aspect. Probably, because it does not belong to the challenges that they are facing in their daily life. They were from industry or applied research working for industry. Accordingly, their systems thinking was limited to business objectives. Business objectives include Environmental, Social, and corporate Governance (ESG) factors to satisfy shareholders and investors. ESG is a market-based solution and as such, it can probably not adequately address systemic market failures [27][28].

A recent publication from the office of the European Union with the title "Managing the implementation of the SDGs" [29] states that "implementation of the SDGs requires readiness of public sector organizations and political leaders to design and manage effective governance. This requires rethinking of institutions, instruments, skills, human resources development and governance processes on at least the same scale as the rethinking of energy, economy, food and other systems that are essential for living well within the limits of the planet." Both kinds of rethinking relate to our broader scope of engineering. The rethinking of institutions relates to that what we observed in complexity science and sustainability science. It changes "how to get things done" and not only "what should be done".

Safety science and related regulatory science is less fundamental and more related to the "what should ne done".

## *Research addressing dynamics of SoS*

This dynamic adaption of "what should be done" relates to one notion that needs to be considered in SoS engineering but we were mainly referring to the dynamic behavior of smart systems. Smart grid, smart mobility, smart manufacturing, smart city and many other smart systems come along with flexibility, runtime adaptation and dynamism. IEC is curtly working on a definition of smart [30] and the current proposal explicitly refers to adaption: "ability of a system to interpret data…to identify and adapt to changes; and to improve prediction and action". Technical systems shall become more adaptive, but why? In the following, we will reflect on the driver for this dynamic adaption of smart systems and its relation to dynamic adaption of regulation.

**Is sustainability the main driver for dynamics of SoS?** Currently, it is but this was not always the case. The smart systems strongly relate to ubiquitous computing which Mark Weiser introduced in 1988. In his first paper [31] about ubiquitous computing he sketched a future scenario of a luxury life. Paper ends with "Machines that fit the human environment, instead of forcing humans to enter theirs, will make using a computer as refreshing as taking a walk in the woods". It was clearly not about lowering the environmental impact but about increasing comfort by means of technology without mentioning consequences for the environment. Many business ideas in context of industry 4.0 and lot size 1 were also not driven by sustainability. Individualized products and shorter time to market for new products can have consequences for the environment. However, when the term industry 4.0 was introduced in 2011, it was already mentioned that it can contribute to green production, e.g., by measuring $CO_2$ footprint using digital twins. Sustainability was considered from the very beginning of industry 4.0 and is not a reasonable criterion to introduce the term industry 5.0. Currently, sustainability is the main driver for all smart systems.

**How does dynamics of SoS support sustainability?** Dynamic adaption can contribute to the three sustainability strategies efficiency, sufficiency and consistency. Most prominent in our examples was efficiency. A smart grid tries to offer maximal energy with resources that minimize impact on the environment. A smart mobility system tries to provides maximal mobility with resources that minimize impact on the environment and so on. The concept of efficiency is tightly coupled with the concept of sufficiency, because being twice as efficient does not help if consume is tripled. Technical systems can influence consumer behavior. A prominent example is given by social media. It can nudge humans very precisely with the right piece of information at the right point in time. Instead of nudging humans to consume more, it is also possible to nudge them to consume less or behave otherwise sustainable [32]. Dynamic pricing can be a powerful means here even though it is primarily a means of efficiency and optimizing the usage of resources. Last but not least, dynamic adaption can also support consistency. Consistency reconciles nature with technology. For instance, field robots do mechanical weeding and avoid thus the usage of chemical weed. They have negligible soil compaction and adapt their path so that fields can have wildflower verge. If we compare the three different use cases of dynamic behavior, then **efficiency** is most prominent.

**What limits the application of dynamics for efficiency and how to deal with?** In many cases, safety and efficiency are in conflict. Conventional safety follows the Keep it Simple, Stupid (KISS) principle. Worst-case assumptions shall guarantee that the system is safe in all situations. This typically implies that the systems behave not efficient in cases where the worst-case is not present. To exemplify this, we consider the hierarchy of cyber-physical systems from the roadmap "Safety, Security, and Certifiability of Future Man-Machine Systems" of the safetrans association [33] (cf. Figure 2).

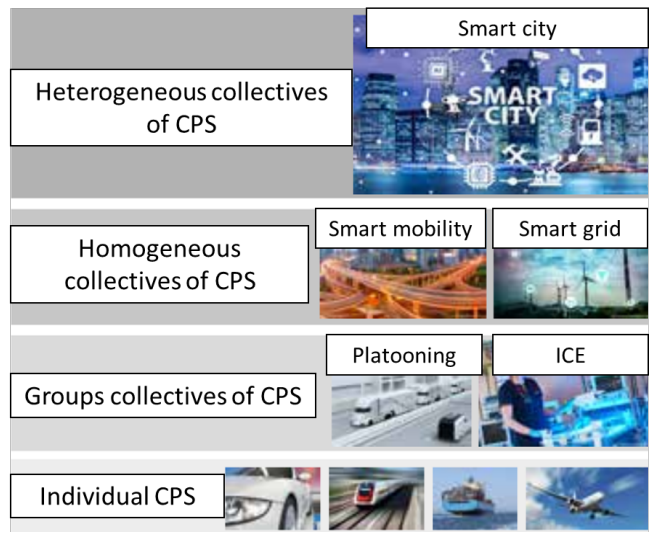

Figure 2 – Hierarchy of CPS from the safetrans roadmap

Prominent examples for SoS that are groups of cyber-physical systems (CPS) are truck platooning and an integrated clinical environment. In truck platooning, trucks build convoys and drive very close behind each other in order to be in the wind shadow and safe fuel. In this example, the conflict between safety and efficiency here is reflected in the distance. A smaller distance is better from an efficiency perspective but a safe distance must be kept in order to avoid rear-end collisions. Worst-case assumptions about factors that influence the safe distance lead to an unnecessary large distance. One approach to minimize such worst-case assumptions is that the systems share safety-critical information in a digital twin. The leader truck can inform the follower truck about its breaking capability, because the follower has to keep a larger distance if it assumes the best possible breaking capability. We can distinguish between two cases. First, the leader sends which breaking capability it has in case of ideal road friction and other influence factors. This information is sent together with other information when the trucks meet in order to evaluate under which conditions they can form a platoon. Second, the leader sends it permanently depending on the current road friction and other current influence factors. This information is permanently evaluated in order to adapt the safety distance. This case is exemplified in [34] by means of digital dependability identities (DDI). A DDI is the entity that carries all the safety-relevant information. It is a machine machine-readable assurance case. An assurance case is a "reasoned, auditable artefact created that supports the contention that its top-level claim (or set of claims) is satisfied, including systematic argumentation and its underlying evidence and explicit assumptions that support the claim(s)" (ISO/IEC/IEEE 15026-3:2022). For instance, if the top-level claim is that the provided value of the breaking capability is correct, then the argumentation would explain why and make calculation assumptions of the algorithms that determine the value explicit. DDIs implement the ideas in [35] and [36] in order to deal with the open context or the operational independence of constituent systems. It relates mainly to the dynamic that constituent systems can enter and leave a SoS.

In our examples of a smart grid or smart mobility, we also discussed another kind of dynamic. This kind refers to the intelligent, flexible usage of resources/CS in order to fulfill the mission of the SoS. It is more related to resilience, emergence and complexity. The loss events are typically not directly life threatening but nonetheless severe and can cause indirectly physical injury. An example here is a large-scale power outage. The loss events occur, because (emergent) SoS behavior generates a hazardous condition instead of keeping an appropriate distance to such conditions. Accordingly, safety concepts have to influence the SoS behavior to maintain a safe distance to such conditions. An example here is the smart grid traffic light concept [37]. The idea behind the traffic light concept is that for a particular period of time and a particular network segment, the network status can be described using one of the colors, "green", "amber" and "red. The latter two refer to a danger to network stability. This can result from energy trading that takes efficiency into account but abstracts from the physical limits of the network structure. Accordingly, trading rules are dynamically changed at runtime. Depending on the relevant traffic light color, certain rules apply in the respective network segment for the interaction of all relevant market roles such as suppliers, balance responsible parties, generators, storage facility operators and the statutory regulated role of the network operator. The latter statutory regulation has a direct relation to the dynamic behavior of smart grid. For this reason, the dynamism of smart systems goes hand in hand with (dynamic adaption of) regulation. However, dynamism of smart systems is much faster and regulation is often criticized of being too slow with respect to the pace of technological evolution. A promising solution here is to make regulation more goal-based and to establish smart standards [38] that refine regulatory requirements and that can be interpreted by smart systems.

## Summary and Conclusion

We observed that a major driver for SoS engineering is sustainability but often with the sole focus on technology, even though it is well known that governance and regulation are essential for achieving sustainability. From our interviews, workshops and literature studies, we conclude that governance and regulation steer transformation processes while SoS engineering only accelerates the introduction of smart systems that follow this established direction. We use the metaphor of steering and acceleration, because both means for controlling a vehicle and both are not independent. Similarly, engineering of governance and regulation should be aligned with SoS engineering of smart systems. For this reason, we propose to enlarge the scope of SoS engineering. We described this proposal using examples, models and approaches from sustainability science. We also showed that sustainability science; complexity science and safety science are already working in this larger engineering scope. In doing so, they apply systems thinking and transdisciplinary communication. Further, they consider systems of systems. In that sense, SoS engineering already happens in this larger engineering scope. It is only not associated within SoS engineering. At least, we got the impression that the systems engineering community could benefit and contribute more to the work that is done in this field. This motivated us to share the impressions from our research in this paper.

Furthermore, we observed that the SoS characteristics do not include the dynamics that we ascertain in smart systems. Accordingly, SoS engineering does give particular attention to this aspect. We propose to further investigate this aspect and provide means for engineering it with respect to sustainability. We observed that it is mainly used for efficiency and related resilience. We propose to further investigate its usage for sufficiency to avoid that well-known rebound effects of efficiency lower sustainability gains or even lead to sustainability losses. AI-powered nudging has potential here but comes along with many questions like "Is there a business model?" and "What are ethical concerns and how can they be addressed (e.g., with regulation)?". Last but not least, we propose to further investigate and evaluate safety concepts for SoS, because safety is the main barrier for many use cases where SoS can contribute to sustainability.

# Biography

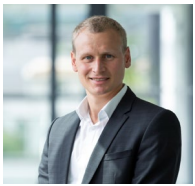

**Rasmus Adler**. Rasmus Adler studied computer science and focused in his PhD on safety-critical adaptive systems. As program manager for autonomous systems and as standardization officer at Fraunhofer IESE, he orchestrates the departments with respect to research about autonomous systems, initiates collaborations with other organizations, and contributes to various standardization activities around dependability of autonomous systems and AI.

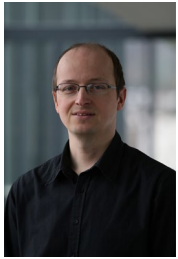

**Frank Elberzhager.** Frank Elberzhager studied computer science and focused in his PhD on software inspections and testing. He is a member in the department "Architecture-Centric Engineering" at Fraunhofer IESE. For more than fifteen years, he has been leading research and industrial projects of different size and criticality. One focus of his current research work relates to digital ecosystems in smart cities and rural areas.

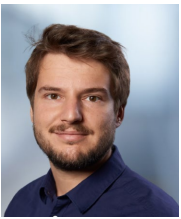

**Florian Balduf.** Florian Balduf studied Industrial Engineering and Management - Focus Mechanical Engineering. He is a member in the department "Embedded Systems Engineering" at Fraunhofer IESE, where he works in the field of model-based systems engineering. In his PhD he focuses on the unification of data models for digital twins in the industry 4.0 domain (AAS), where one focus relates to sustainability information.